\documentclass[a4paper,conference,final]{IEEEtran}
\IEEEoverridecommandlockouts
\usepackage{cite}
\usepackage{amsmath,amssymb,amsfonts}
\usepackage{algorithmic}
\usepackage{graphicx}
\usepackage{textcomp}
\usepackage{subfigure}
\usepackage{stfloats}
\usepackage{cite}
\usepackage{amsmath,amssymb,amsfonts}
\usepackage{algorithmic}
\usepackage{graphicx}
\usepackage{textcomp}
\usepackage{epstopdf}
\usepackage{subfigure}
\usepackage{stfloats}
\usepackage{multirow}
\usepackage{booktabs}
\usepackage{multirow}
\usepackage{graphicx}
\usepackage{epstopdf}
\usepackage{stfloats}
\usepackage{subfigure}
\usepackage{microtype}
\usepackage{setspace}

\usepackage{cite}
\usepackage{color}
\usepackage{amsmath}
\usepackage{array}
\usepackage{pifont}

\usepackage{color}
\usepackage{graphicx}

\def\BibTeX{{\rm B\kern-.05em{\sc i\kern-.025em b}\kern-.08em
    T\kern-.1667em\lower.7ex\hbox{E}\kern-.125emX}}
\DeclareRobustCommand*{\IEEEauthorrefmark}[1]{%
    \raisebox{0pt}[0pt][0pt]{\textsuperscript{\footnotesize\ensuremath{#1}}}}

\begin{document}
\title{Effects of Digital Map on the RT-based Channel Model for UAV mmWave Communications \\\begin{large}\emph{(Invited Paper)} \end{large} }
\author{
\IEEEauthorblockN{Qiuming~Zhu\IEEEauthorrefmark{1,}\IEEEauthorrefmark{*}, Shan~Jiang\IEEEauthorrefmark{1},
Cheng-Xiang~Wang\IEEEauthorrefmark{2,}\IEEEauthorrefmark{3,}\IEEEauthorrefmark{*}, Boyu~Hua\IEEEauthorrefmark{1}, Kai~Mao\IEEEauthorrefmark{1}, Xiaomin~Chen\IEEEauthorrefmark{1}, Weizhi~Zhong\IEEEauthorrefmark{1}}

\IEEEauthorblockA{\IEEEauthorrefmark{1}The Key Laboratory of Dynamic Cognitive System of Electromagnetic Spectrum Space, \\
College of Electronic and Information Engineering, \\
Nanjing University of Aeronautics and Astronautics, Nanjing 211106, China}
\IEEEauthorblockA{\IEEEauthorrefmark{2}National Mobile Communications Research Laboratory, School of Information Science and Engineering, \\
Southeast University, Nanjing 210096, China}
\IEEEauthorblockA{\IEEEauthorrefmark{3}Purple Mountain Laboratories, Nanjing 211111, China}
\IEEEauthorblockA{\IEEEauthorrefmark{*}Corresponding author}
Email: \{zhuqiuming, js0615\}@nuaa.edu.cn, chxwang@seu.edu.cn,\\\{byhua,maokai,chenxm402\}@nuaa.edu.cn}

\maketitle

\begin{abstract}
\par Based on the geometry and ray tracing (RT) theory, a millimeter wave (mmWave) channel model and parameter computation method for unmanned aerial vehicle (UAV) assisted air-to-ground (A2G) communications are proposed in this paper. In order to speed up the parameter calculation, a reconstruction process of scene database on the original digital map is developed. Moreover, the effects of reconstruction accuracy on the channel parameter and characteristic are analyzed by extensive simulations at 28 GHz under the campus scene. The simulation and analysis results show that the simplified database can save up to 50$\%$ time consumption. However, the difference of statistical properties is slight in the campus scenario.
\end{abstract}


%
\IEEEpeerreviewmaketitle
\section{Introduction}
\par UAVs have attracted significant attention from various fields, such as aerial detection, weather monitoring, and agricultural management, due to their high mobility, low cost, and easy deployment \cite{Hayat16}. The mmWave communication technologies can integrate large number of antennas into a small size and provide multi-gigabit transmit ability \cite{Rappaport15, Zhong19}. For the upcoming fifth and beyond fifth generation (5G/B5G) communication systems, the UAV-assisted mmWave communications have been considered an important application scenario, such as the aerial base station and flying relay\cite{Bin19, Wang18, Fan17, Zhang19, Xiao16, Kong17, wu18}. Compared with traditional  mobile channels \cite{Zhu18trans, Bi19, Liu20, Bai19}, UAV channels have some unique properties, i.e., 3D flight, 3D non-stationary propagation environment, and valid scatterers only around ground station (GS) \cite{Cheng18, Zhu18china}. It is essential to explore and understand these unique properties, which are important for optimizing and evaluating the UAV-assisted mmWave communication systems \cite{Zhu19WC, Miao18, Khosravi18}.
\par There are limited theoretical studies or measurements in the literature on UAV channel modeling, especially for the mmWave band.  By upgrading the traditional geometry-based stochastic model (GBSM), several sub-6GHz GBSMs including new properties of UAV communication scenarios were studied in \cite{Cheng18, Zhu18china, Zhu19, Chang19, Cui19}. Several measured results for A2G channels, i.e., path loss, delay spread, and propagation angle can be addressed in \cite{Cui19, Matolak17, Khawaja17}, but only the measurement campaign in \cite{Khawaja17} was designed for the mmWave band.

\par Recently, the RT technique has been adopted as an alternative method of field measurement to aid the mmWave channel modeling. Based on the 3D UAV GBSM and RT method, the authors in \cite{Cheng19} explored the channel parameters at 28GHz in the campus scenario. The temporal and spatial characteristics of A2G channel at 28 GHz in different environmental scenarios were studied in \cite{Khawaja18}. The authors in \cite{Perez19} analyzed the mmWave propagation characteristics at 30 and 60 GHz frequency bands above a Manhattan-like environment. Moreover, the propagation characteristics in mountain terrain and suburban environments were analyzed by RT techniques in \cite{Cui19_ICC, Chu18}.

\par It should be mentioned that the RT method has high complexity and is very time-consuming. Several ray tracing acceleration techniques according to the character of terrain models, i.e., a preliminary selection of the rays and optimization of rectangular meshes, can be addressed in \cite{Degli05}. Since it's unacceptable to run the RT method on the realistic digital map, another easy way is to use a simplified digital map. However, to the best of our knowledge, the effects of digital map accuracy on the UAV mmWave channel characteristics have not been studied yet. This paper intends to fill this research gap. The main contributions and innovations of this paper are generalized as follows:

\par1) A geometry and RT based 3D channel model for UAV A2G mmWave communications is proposed. The model considers the key factors of 3D propagation environment and 3D arbitrary trajectories.

\par 2) A geometric map based computation method for geometric parameters and channel parameters is developed. In order to reduce the computation complexity, a detailed reconstruction process of scene database is also developed. The reconstructed database has a flexible accuracy due to the user defined requirement.

\par 3) By generating three campus scene databases with different accuracies, channel parameters, i.e., power, delay, and angles of different rays, and statistical properties such as the distributions of power spread, delay spread, and angle offset are analyzed and compared for six different trajectories at altitude of 75m.

\par The rest of this paper is arranged as follows. In Section II, a 3D UAV mmWave channel model based on RT principle is proposed. The map-based computation method for channel parameters as well as the reconstruction process of scene database are presented in Section III. Extensive simulations are conducted with different data scenes and simulation results are analyzed in Section IV. At last, conclusions are given in Section V.

%
%
%
%


\section{UAV-assisted mmWave Channel Model}
\begin{figure}[!b]
 {\includegraphics[width=80mm,height=58mm]{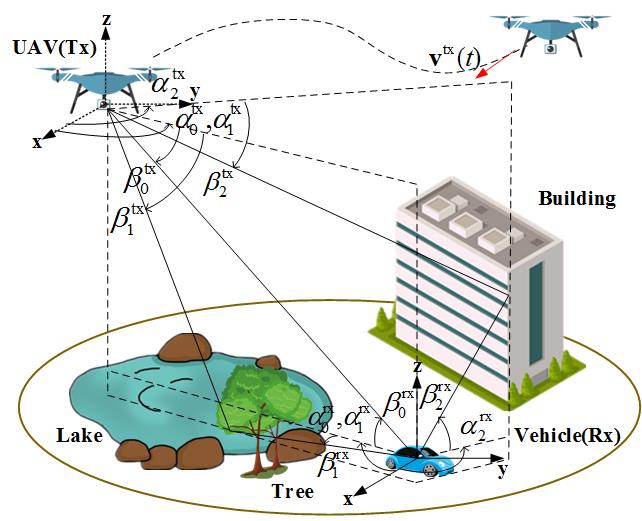}}
\caption{UAV-assisted A2G mmWave communication scenario.}
\label{fig:1}
\end{figure}
\par Fig. 1 shows a typical UAV to ground mmWave communication downlink, where the UAV and vehicle are the transmitter (Tx) and receiver (Rx), respectively. Since the UAV flies fast, the azimuth angle of departure (AAoD) $\alpha _l^{{\rm{tx}}}(t)$, elevation angle of departure (EAoD) $\beta _l^{{\rm{tx}}}(t)$, azimuth angle of arrival (AAoA) $\alpha _l^{{\rm{rx}}}(t)$, and elevation angle of arrival (EAoA) $\beta _l^{{\rm{rx}}}(t)$ are time-variant and related with the particular locations of Tx, Rx, and scatterers. In this paper, combining the RT principle and dynamic propagation scenario, the UAV A2G channel is represented as the sum of different rays as
\setcounter{equation}{0}
\begin{align}
h(t,\tau ,\alpha ,\beta ) &= \sum\limits_{l = 0}^{L(t)} {{P_l}(t)} {{\rm e}^{\frac{{2{\rm{\pi }}}}{{{\lambda _0}}}\int_0^t {{{\left( {{{\bf{v}}^{{\rm{tx}}}}(t)} \right)}^{\rm{T}}}{\bf{r}}_{_l}^{{\rm{rx}}}(t)} {\rm{d}}t + {\psi _l}}}\notag\\
  &\cdot \delta (t - {\tau _l}(t))\delta (\alpha _{}^{{\rm{tx}}} - \alpha _l^{{\rm{tx}}}(t))\tag{1}\\
  &\cdot \delta (\beta _{}^{{\rm{tx}}} - \beta _l^{{\rm{tx}}}(t))\delta (\alpha _{}^{{\rm{rx}}} - \alpha _l^{{\rm{rx}}}(t))\notag\\
  &\cdot \delta (\beta _{}^{{\rm{rx}}} - \beta _l^{{\rm{rx}}}(t))\notag
\end{align}
\noindent where $L(t)$ is the number of valid rays, ${{\rm{P}}_l}{\rm{(t)}}$, ${\psi _l}$, and ${\tau _l}(t)$ denote the power gain, random initial phase, and delay of $l$th ray, respectively, and $\lambda_0 $ denotes the wavelength. Note that $l = 0$ denotes the light-of-sight (LoS) ray, otherwise the non-light-of-sight (NLoS) ray. In (1), ${{\bf{V}}^{{\rm{tx}}}}(t)$ means the UAV velocity, ${\bf{r}}_l^{\rm tx/rx}(t)$ means the directional unit vector of the receiving and transmitting signals of the $l$th ray. Since the spherical unit vector function in the 3D space is defined as
\setcounter{equation}{1}
\begin{equation}
\begin{array}{l}
{\bf{s}}(v(t),u(t)) = \left[ {\begin{array}{*{20}{c}}
{\cos v(t)\cos u(t)}\\
{\cos v(t)\sin u(t)}\\
{\sin v(t)}
\end{array}} \right]
\label{2}
\end{array}
\end{equation}
\noindent where $v(t)$ and $u(t)$ denote the azimuth angle and elevation angle, respectively, ${{\bf{V}}^{{\rm{tx}}}}(t)$ and ${\bf{r}}_l^{{\rm{tx/rx}}}(t)$ can be obtained by $\left\| {{v^{{\rm{tx}}}}} \right\|{\bf{s}}\left( {\beta _{}^v(t),\alpha _{}^v(t)} \right)$ and ${\bf{s}}\left( {\beta _l^{{\rm{tx/rx}}}(t),\alpha _l^{{\rm{tx/rx}}}(t)} \right)$, respectively, $\alpha _{}^v(t)$ and $\beta _{}^v(t)$ represent the azimuth and elevation angles of movement, respectively.

\section{Map-based Computation Method for Channel Parameters}
\subsection{Flowchart of Parameter Computation}
\par The map-based computation method in this paper aims to calculate the channel parameters, according to the locations of transceivers and their propagation environments. The RT method is based on the geometrical theorem of diffraction, uniform theory of diffraction (UTD), and field intensity superposition principle. Due to its complexity and time consumption, it's more applicable to run the RT method on a simplified digital map. Thus, the proposed algorithm in this paper mainly includes two steps, i.e., reconstruction of scene data and parameter computation by RT techniques, where the reconstruction is very important and flexible to control the complexity and accuracy. Based on the reconstructed database, RT techniques are adopted to track the electromagnetic propagation process, i.e., direction, reflection, and diffraction. Parameters such as power, delay, phase, and angle of each ray can be obtained.

\subsection{Reconstruction Process of Scene Database}
\par The original geometric map of interesting area can be obtained from the Google earth or other similar softwares. It includes numerous building elements, e.g., internal yards, vegetation, roof structures, etc., and the complex terrain, which can result in several hours of computation time.

\par In this paper, we obtain the original map in the form of digital terrain model (DEM), which has the detailed information of latitude, longitude and elevation of all the points. These information can be used to reconstruct many regular or irregular triangle facets to approximately describe the surface of terrain. When considering the large terrain with buildings, the triangles will no longer be regular ones with similar size in order to connect with the building. Fig. 2 (a) shows the original map of Nanjing University of Aeronautics and Astronautics (NUAA) campus, and Fig. 2 (b) and Fig. 2 (c) denote the reconstructed scenes of terrain and terrain with buildings, respectively.
\begin{figure}[h]
	\centering
	\subfigure[]{
		\begin{minipage}[t]{0.333\linewidth}
			\centering
			\includegraphics[width=0.95in]{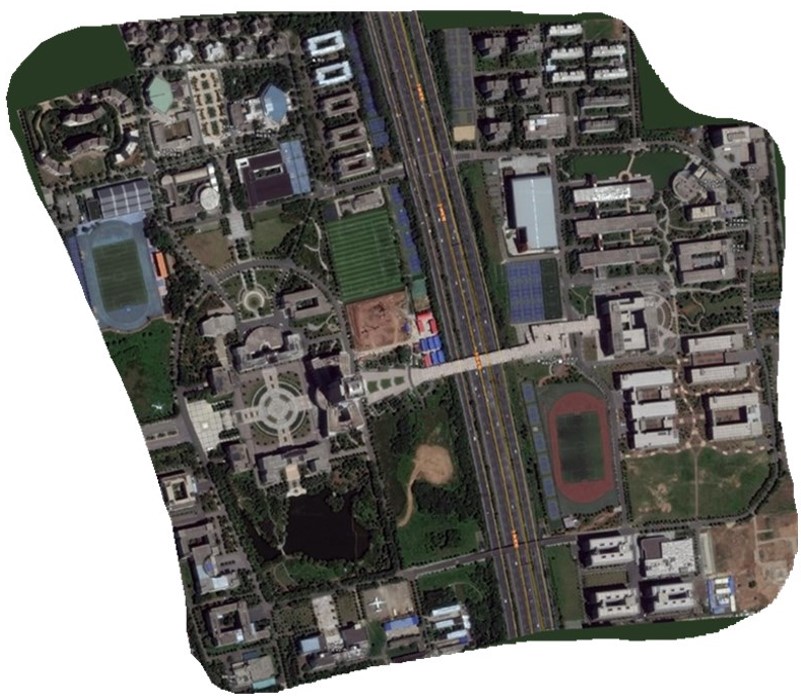}
		\end{minipage}%
	}%
	\subfigure[]{
		\begin{minipage}[t]{0.333\linewidth}
			\centering
			\includegraphics[width=0.95in]{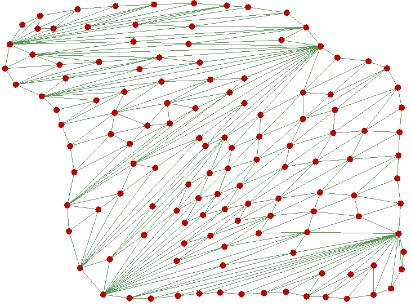}
		\end{minipage}%
	}%
	\subfigure[]{
		\begin{minipage}[t]{0.333\linewidth}
			\centering
			\includegraphics[width=0.95in]{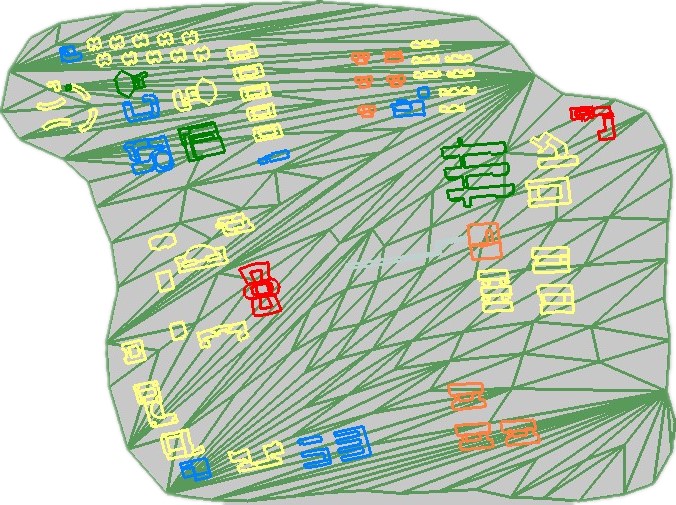}
		\end{minipage}
	}
	\centering
	\caption{(a) Original map and reconstructed database of (b) terrain and (c) terrain with buildings of NUAA campus}
\end{figure}

\par The impacts of structures and materials can be assigned according to the reality. For simplicity, we consider the building material of all buildings as concrete. Foliage area consisting of branches and leaves of trees are also considered. Other terrain covers such as lamp and sign boards are ignored due to their smaller size.

\subsection{Computation of Geometric Parameters}
\par It's assumed that the vehicle on the ground is relatively stationary and taken as the origin of coordinates. The 3D location vector of the $l$th scatterer is assumed to be time-invariant during a short analytical period as
\setcounter{equation}{2}
\begin{equation}
{\bf{L}}_l^{\rm{S}}(t) = D_l^{\rm{rx,S}}(t) \cdot {\bf{r}}_l^{{\rm{rx,S}}}(t)
\label{3}
\end{equation}
\noindent where $D_l^{\rm{rx,S}}(t)$ denotes the distance between the vehicle and the $l$th scatterer. Similarly, the location vector of UAV at the initial moment is
\setcounter{equation}{3}
\begin{equation}
\begin{array}{l}
{{\bf{L}}^{{\rm{tx}}}}\left( {{t_0}} \right) = D_0^{{\rm{tx}}}\left( {{t_0}} \right) \cdot {\bf{r}}_0^{{\rm{tx}}}\left( {{t_0}} \right)
\label{4}
\end{array}
\end{equation}
\noindent where $D_0^{{\rm{tx}}}\left( {{t_0}} \right)$ denotes the distance of two terminals at the initial moment. Therefore, the time-varying position vector of UAV denoted as ${{\bf{L}}^{{\rm{tx}}}}(t)$ at the time $t$ can be expressed as
\setcounter{equation}{4}
\begin{equation}
\begin{array}{l}
{{\bf{L}}^{{\rm{tx}}}}(t) = {{\bf{L}}^{{\rm{tx}}}}(t - \Delta t) + {{\bf{v}}^{\rm{tx}}}(t) \cdot \Delta t
\label{5}
\end{array}
\end{equation}
\noindent where $\Delta t$ denotes the time lag.
\par Based on the real-time locations of the vehicle, UAV, and scatterers, the Euclidean metric between the vehicle and UAV, the scatterer and UAV can be updated by (6) and (7), respectively, where $D_l^{{\rm{S,tx}}}(t)$ denotes the distance between the UAV and the $l$th scatter. Therefore, the time-varying delay of the $l$th NLoS ray and LoS ray at the time $t$ can be obtained by dividing the speed of light $c$, respectively.
\begin{figure*}[!tb]
\normalsize
\setcounter{equation}{5}
\begin{equation}
\begin{array}{l}
{D^{{\rm{tx}}}}(t) = \left\| {{{\bf{L}}^{{\rm{tx}}}}(t - \Delta t) + {{\bf{v}}^{{\rm{tx}}}}(t) \cdot \Delta t} \right\|\\
 = \sqrt{\begin{array}{l}
{\rm{ }}{D^{{\rm{tx}}}}{(t - \Delta t)^2} + {\left( {\left\| {{{\bf{v}}^{{\rm{tx}}}}(t)} \right\|\Delta t} \right)^2} + 2{D^{{\rm{tx}}}}(t - \Delta t)\left\| {{{\bf{v}}^{{\rm{tx}}}}(t)} \right\|\Delta t \cdot \\
\left( \begin{array}{l}
\cos \beta _0^{{\rm{rx}}}(t - \Delta t)\cos{\alpha ^v}(t - \Delta t)\cos \left( {\alpha _{\rm{0}}^{{\rm{rx}}}(t - \Delta t) - {\beta ^v}(t - \Delta t)} \right)\\
 + \sin \beta _{\rm{0}}^{{\rm{rx}}}(t - \Delta t)\sin{\beta ^v}(t - \Delta t)
\end{array} \right)
\end{array}}
\label{6}
\end{array}
\end{equation}
\end{figure*}
\begin{figure*}[!tb]
\normalsize
\setcounter{equation}{6}
\begin{equation}
\begin{array}{l}
D_l^{{\rm{S,tx}}}(t) = \left\| {{{\bf{L}}^{\rm{tx}}}(t) - {\bf{L}}_l^{\rm{S}}} \right\|\\
 = \sqrt {\begin{array}{l}
{\left( {D_{}^{{\rm{tx}}}(t - \Delta t) - D_l^{{\rm{rx}}}} \right)^2} + {\left( {\left\| {{\bf{v}}_{}^{{\rm{tx}}}(t)} \right\|\Delta t} \right)^2} + 2\left( {D_{}^{{\rm{tx}}}(t - \Delta t) - D_l^{{\rm{rx}}}} \right)\left\| {{\bf{v}}_{}^{{\rm{tx}}}(t)} \right\|\Delta t\\
 \cdot \left( \begin{array}{l}
\cos \beta _l^{{\rm{rx}}}(t - \Delta t)\cos\alpha _{}^v(t - \Delta t)\cos \left( {\alpha _l^{{\rm{rx}}}(t - \Delta t) - \beta _{}^v(t - \Delta t)} \right)\\
 + \sin \beta _l^{{\rm{rx}}}(t - \Delta t)\sin{\beta _v}(t - \Delta t)
\end{array} \right)
\end{array}}
\label{7}
\end{array}
\end{equation}
\end{figure*}
\par Let us set ${\bf{L}}_{l,x}^{{\rm{tx/rx/S}}}(t),{\bf{L}}_{l,y}^{{\rm{tx/rx/S}}}(t),{\bf{L}}_{l,z}^{{\rm{tx/rx/S}}}(t)$ as the components in x, y, and z direction of the 3D locations of UAV, vehicle and scatterer, respectively. Then, we convert the Cartesian coordination into the spherical coordination. The time-variant EAoD, EAoA and AAoD, AAoA for the $l$th NLoS ray can be calculated by (8) and (9), respectively.
\begin{figure*}[!tb]
\normalsize
\setcounter{equation}{7}
\begin{equation}
\beta _l^{{\rm{tx/rx}}}(t) = \arctan (\frac{{{\bf{L}}_{l,z}^{\rm{S}}(t) - {\bf{L}}_{l,z}^{{\rm{tx/rx}}}(t)}}{{\sqrt {{{({\bf{L}}_{l,x}^{\rm{S}}(t) - {\bf{L}}_{l,x}^{{\rm{tx/rx}}}(t))}^2} + {{({\bf{L}}_{l,y}^{\rm{S}}(t) - {\bf{L}}_{l,y}^{{\rm{tx/rx}}}(t))}^2}} }})
\label{8}
\end{equation}
\end{figure*}
\begin{figure*}[!tb]
\normalsize
\setcounter{equation}{8}
\begin{equation}
\alpha _l^{{\rm{tx/rx}}}(t) = \arcsin\left( {\frac{{{\bf{L}}_{l,y}^{\rm{S}}(t) - {\bf{L}}_{l,y}^{{\rm{tx/rx}}}(t)}}{{\sqrt {{{({\bf{L}}_{l,x}^{\rm{S}}(t) - {\bf{L}}_{l,x}^{{\rm{tx/rx}}}(t))}^2} + {{({\bf{L}}_{l,y}^{\rm{S}}(t) - {\bf{L}}_{l,y}^{{\rm{tx/rx}}}(t))}^2}} }})} \right)
\label{9}
\end{equation}
\end{figure*}

\subsection{Computation of Channel Parameters}
\par Since the propagation condition of LoS ray in the UAV to ground communications is similar to the free space environment, the power gain of LoS ray, defined by the ratio of transmitting power and receiving power in dB, can be calculated by
\setcounter{equation}{9}
\begin{equation}
    {P_{\rm{0}}}(t) = 32.44 + 20{\rm{lo}}{{\rm{g}}_{10}}(f) + 20{\rm{lo}}{{\rm{g}}_{10}}(D_{\rm{0}}^{{\rm{tx}}}(t))
\label{10}
\end{equation}
\noindent where $f$ denotes the carrier frequency in MHz.

\par For the case of NLoS rays, the reflection ray and diffraction ray should be considered separately. Firstly, the electric field intensity of LoS condition can be expressed as
\setcounter{equation}{10}
\begin{equation}
{{\bf{E}}_0}(t) = {\bf{E}}_{}^{\rm 1m}\frac{{{{\rm e}^{ - jkD_{\rm{0}}^{{\rm{tx}}}(t)}}}}{{D_{\rm{0}}^{{\rm{tx}}}(t)}}
\label{11}
\end{equation}
\noindent where ${\bf{E}}_{}^{\rm 1m}$ is the electric field intensity 1 m away from the UAV, and $k$ is the number of waves. Moreover, the electric field intensity of reflected ray can be described as

\setcounter{equation}{11}
\begin{equation}
{\bf{E}}_l^{\rm{R}}(t) = {{\bf{E}}_0}R\frac{{{{\rm e}^{ - lk(D_l^{{\rm{rx,S}}} + D_l^{{\rm{S,tx}}}(t))}}}}{{D_l^{{\rm{rx,S}}} + D_l^{{\rm{S,tx}}}(t)}}
\label{12}
\end{equation}
\noindent where $R$ is the reflection coefficient with respect of the polarization mode, i.e., horizontal or vertical, and can be calculated by
\setcounter{equation}{12}
\begin{equation}
{R_\parallel } = \frac{{{\varepsilon _{\rm r}}\cos \theta  - \sqrt {{\varepsilon _{\rm r}} - {{\sin }^2}\theta } }}{{{\varepsilon _{\rm r}}\cos \theta  + \sqrt {{\varepsilon _{\rm r}} - {{\sin }^2}\theta } }}
\label{13}
\end{equation}
\noindent and
\setcounter{equation}{13}
\begin{equation}
{R_ \bot } = \frac{{\cos \theta  - \sqrt {{\varepsilon _{\rm r}} - {{\sin }^2}\theta } }}{{\cos \theta  + \sqrt {{\varepsilon _{\rm r}} - {{\sin }^2}\theta } }}
\label{14}
\end{equation}
\noindent where ${\varepsilon _{\rm r}}$ is relative permittivity and $\theta $ is incident angle.

\par According to the UTD theory, the electric field intensity of diffracted ray can be expressed as
\begin{equation}
    \begin{split}
     {\bf{E}}_l^{\rm{D}}(t) &= \frac{{{{\bf{E}}_0}}}{{D_l^{{\rm{rx,S}}}(t)}}I\sqrt {\frac{{D_l^{{\rm{rx,S}}}(t)}}{{D_l^{{\rm{S,tx}}}(t)(D_l^{{\rm{rx,S}}}(t) + D_l^{{\rm{S,tx}}}(t))}}} \\
      &\cdot{{\rm e}^{ - jk(D_l^{{\rm{rx,S}}}(t) + D_l^{{\rm{S,tx}}}(t))}}
    \end{split}
\end{equation}
\noindent where $I$ is diffraction coefficient and can be expressed as
\setcounter{equation}{15}
\begin{equation}
I(t) = {I_1}(t) + {I_2}(t) + R_0^{}{I_3}(t) + R_n^{}{I_4}(t)
\label{16}
\end{equation}
\noindent where ${R_0}$ and ${R_n}$ are the reflection coefficients for the zero and $n$ face, respectively, the component ${I_i}(t)(i = 1,2,3,4)$ can be further calculated by
\begin{equation}
    \begin{split}
     {I_i}(t) &= \frac{{ - {{\rm e}^{ - j\pi /4}}}}{{2n\sqrt {2\pi k} }}{\rm ctg}{\gamma _i}\\
      &\cdot {F_0}(2k\frac{{D_l^{{\rm{rx,S}}}(t)D_l^{{\rm{S,tx}}}(t)}}{{(D_l^{{\rm{rx,S}}}(t) + D_l^{{\rm{S,tx}}}(t))}}{n^2}{\sin ^2}{\gamma _i})
    \end{split}
\end{equation}
\noindent where $n\pi$ is exterior angle of the wedge, ${F_0}$ is transition function, and ${\gamma _i}$ is given by
\begin{equation}
    \begin{split}
      {\gamma _1}&= \left[ {\pi  - (\vartheta  - \theta )} \right]/2n,{\gamma _2} = \left[ {\pi  + (\vartheta  - \theta )} \right]/2n\\
      {\gamma _3}&= \left[ {\pi  - (\vartheta  - \theta )} \right]/2n, {\gamma _4} = \left[ {\pi  + (\vartheta  - \theta )} \right]/2n\\
    \end{split}
\end{equation}
\noindent where $n\pi  - \vartheta $ is reflection angle.

\par Finally, the power gain of the $l$th NLoS ray can be obtained by adding a certain extra loss to the LoS condition in dB as
\setcounter{equation}{18}
\begin{equation}
{P_l}(t) = {P_0}(t) + L_l^{}(t){\rm{,}}\;l \ne 0
\label{19}
\end{equation}
\noindent where $L_l^{}(t)$ is related with the propagation condition and can be calculated by
\setcounter{equation}{19}
\begin{equation}
L_l^{}(t) = 20{\log _{10}}\left| {\frac{{{\bf{E}}_l^{{\rm{R/D}}}(t)}}{{{\bf{E}}_0^{}(t)}}} \right|,\;l \ne 0.
\label{20}
\end{equation}

\section{Simulation Results and Analyses}
\subsection{Scene Database Setup}
\par The analyzing scenario is the campus of NUAA, which is a typical open area with buildings, lakes, vegetation, trees, and viaduct. The main area contains 66 buildings with an average height of about 30 m, and the open ground is mostly wet soil.
\par In order to illustrate the effect of database accuracy on the channel parameters and statistical properties, the scene database is reconstructed from the original map with three different criteria as shown in Fig. 3. In this figure, all obstacles with different heights are shown with different colors. Database I only contains high buildings above 20~m. Database II contains all buildings above 5 m, while Database III includes all buildings as well as vegetation and lakes. Table I describes the major simulation parameters, and the UAV and vehicle are all equipped with vertically polarized omnidirectional antennas.
\begin{figure}[!tb]
\centering
 {\includegraphics[width=80mm,height=55mm]{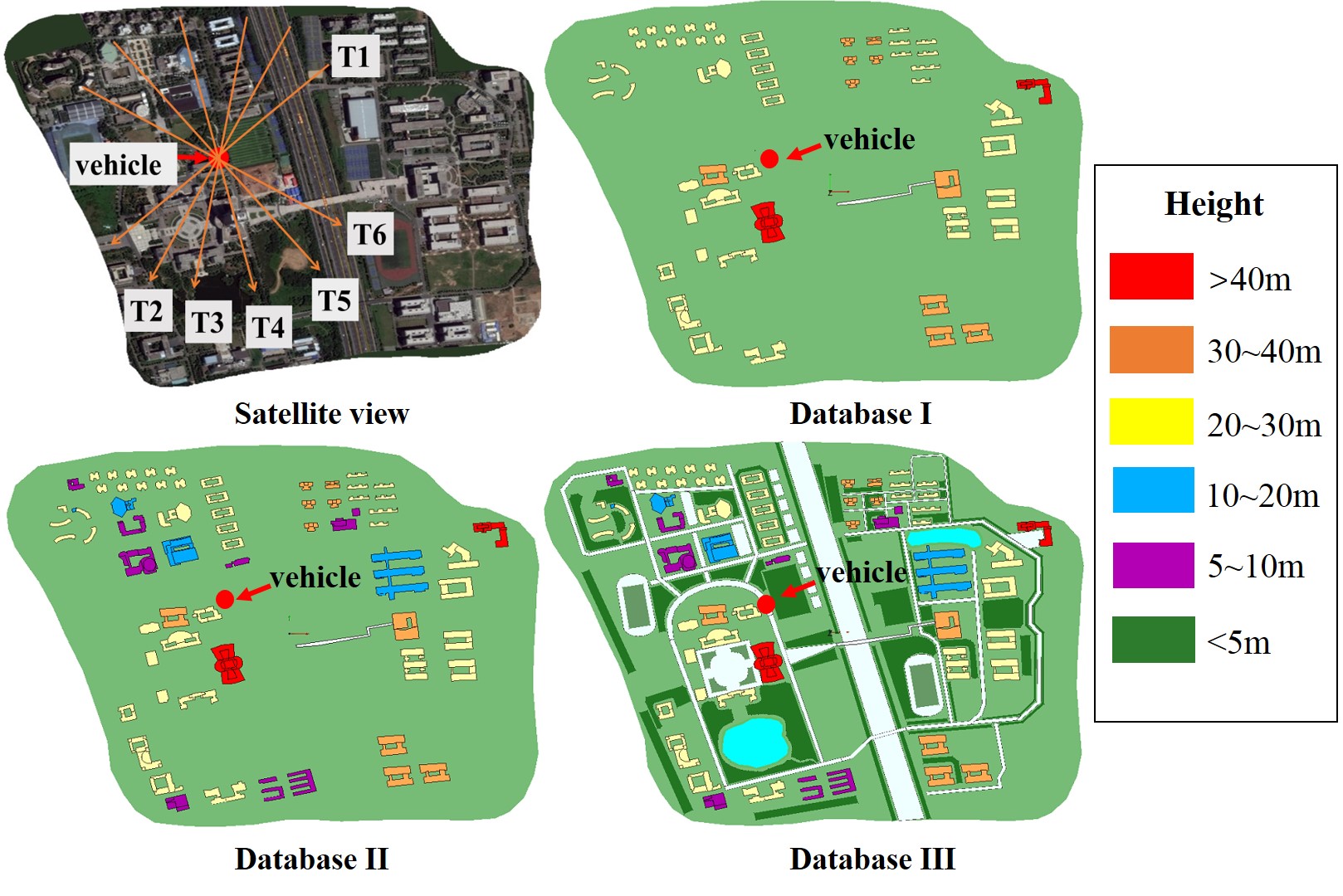}}
\caption{Reconstructed database with different accuracies.}
\label{fig:3}
\end{figure}

\begin{table}[!tb]
\renewcommand{\arraystretch}{1.4}
\caption{Parameters of RT simulations.}
\label{table I}
\centering
\begin{tabular}{|p{3cm}|p{3cm}|}
\hline
Parameter & Value \\ \hline
Frequency & 28 GHz \\ \hline
Bandwidth & 500 MHz \\\hline
Transmitting power & 20 dBm \\\hline
Antenna type & omnidirectional \\\hline
UAV height & 75 m \\\hline
Vehicle height & 2 m \\\hline
UAV speed & 10 m/s \\\hline
Duration & 100 s \\\hline
\end{tabular}
\end{table}

\subsection{Comparison and Analysis}
\par The simulation hardware platform is Inter (R) Xeon (R) E5--1630 with the CPU frequency 3. 7 GHz, and the memory is 16.0 G. We performed many simulations and the average simulation time of three databases is 23.67 s, 33.67 s, and 46.67 s respectively. The database I saves up 50 $\%$ time consumption compared with database III. In order to obtain the averaged statistical properties, six trajectories at the altitude of 75 m are selected as shown in Fig. 3. For these six trajectories, the LoS ray is always existing and the distance between the UAV and vehicle at the same time instant, i.e., totally 100 discrete time instants or spatial samples, are the same. In the simulation, the rays with the power below -45 dB compared with the LOS ray are abandoned. The number of valid rays and their power gains are tracked. The averaged number of valid rays for six trajectories over time and the corresponding distribution of relative power gain under different database are performed and compared in Fig. 4. Here, the relative power gain means the gain difference between the gain of each NLoS ray with the one of LoS ray, and it should be in the range of -45 dB $\sim$ 0 dB. The simulation results show that the number of valid rays under database III is the largest due to its rich scattering environment. Moreover, the number under Database III is 0 $\sim$ 7 larger than the one under Database I and 0 $\sim$ 4 over the one under Database II. The averaged offset values are 2.35 and 1.64, respectively.

\begin{figure}[tb]
	\centerline{\includegraphics[width=0.5\textwidth]{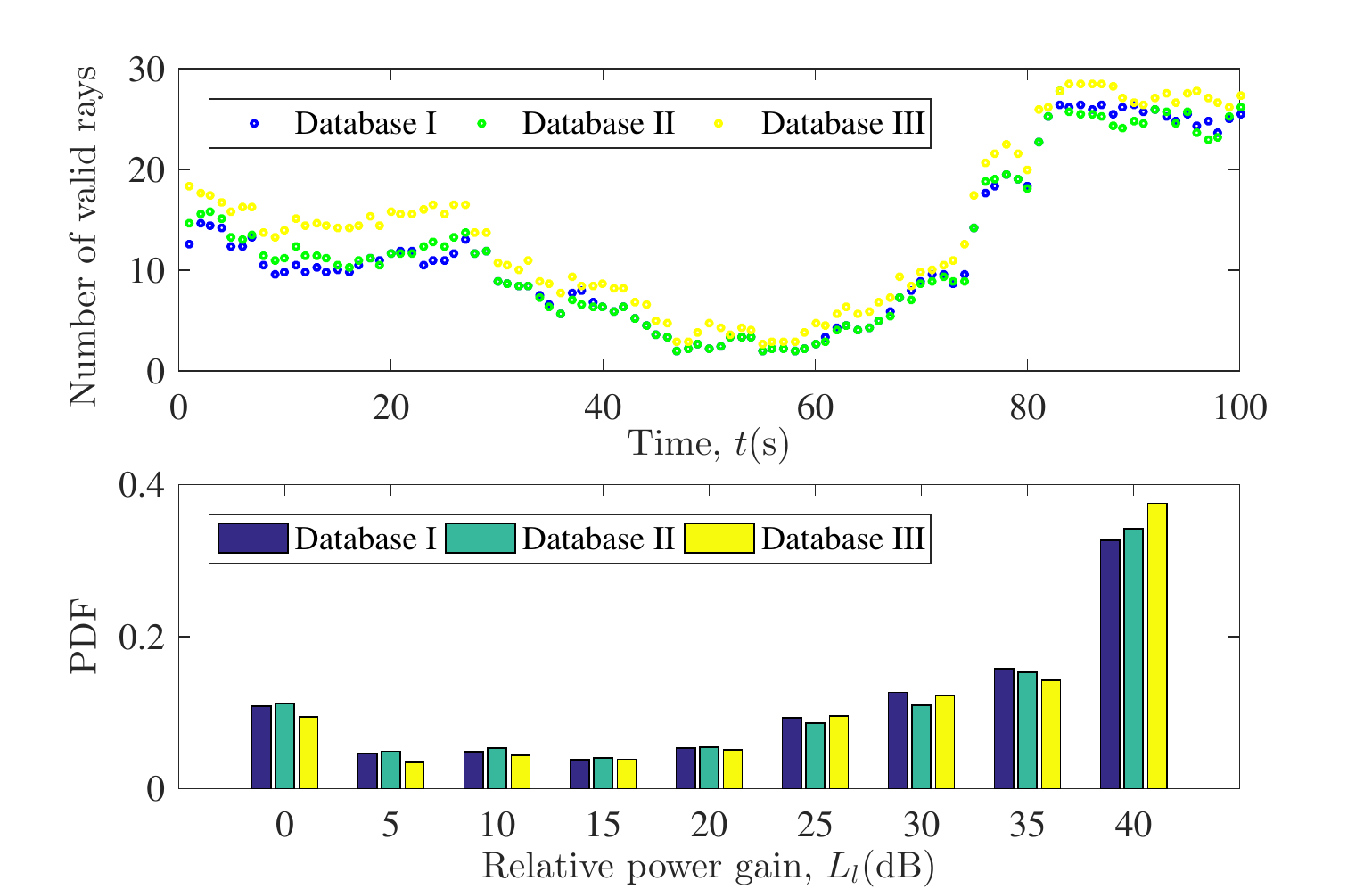}}
	\caption{The number of valid rays over time and distribution of relative power gain.}
	\label{fig:4}
\end{figure}

%
%

\par The delay and delay spread, i.e., the power-weighted root mean square value of delay, are important for evaluating the quality of fading channels. The delays of all rays and the corresponding delay spreads are also calculated. The distributions are statistically analyzed and compared in Fig. 5. In the figure, the relative delay is adopted instead of the absolute delay in order to exclude the effect of time-variant LoS delay due to the UAV's position. The delay spread in Fig. 5 shows that it ranges from 0 ns to 700 ns and mainly concentrates between 0 ns and 100 ns, which is consistent with the measured results. Note that the main distributions of delay spread under three databases are similar, but the  distribution of relative delay under Database I is significantly different with others.
\begin{figure}[tb]
	\centerline{\includegraphics[width=0.5\textwidth]{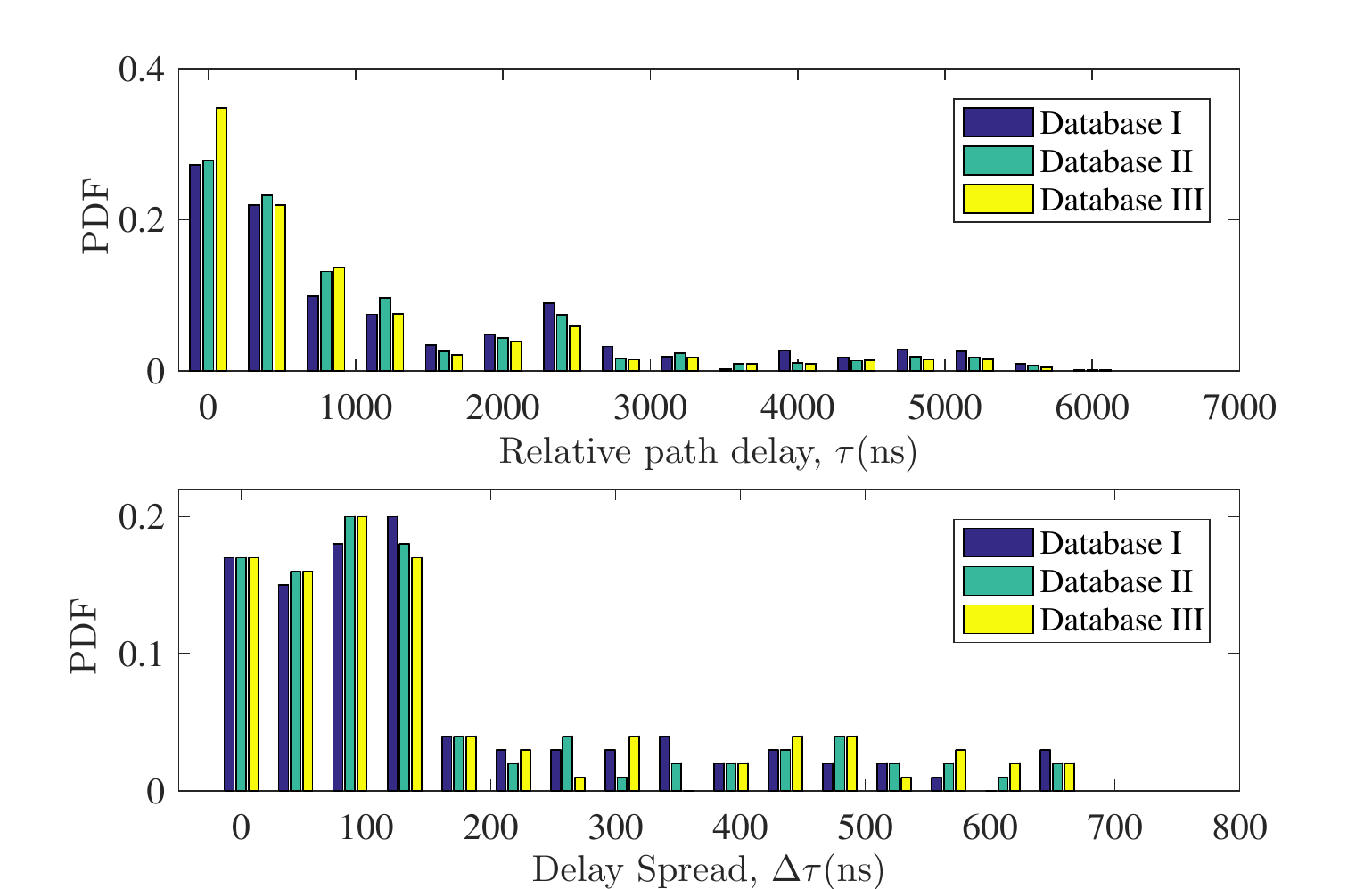}}
	\caption{Statistical distributions of relative path delay and DS.}
	\label{fig:5}
\end{figure}

\par In the simulation, we also record the traces of AAoA and EAoA under different databases and compared in Fig.6. In the figure, the angles of LoS ray are shown in red, which clearly shows that the elevation angle changes smaller than the azimuth angle. The reason is that the UAV¡¯s altitude during the flight remains unchanged, while the azimuth angle changes greatly as the UAV¡¯s position changes. Furthermore, we calculate the angle offset of each ray by subtracting the angle of LoS ray and give the distributions of AAoD, EAoD, AAoA, and EAoA in Fig. 7. We find that the lognormal distribution can fit the simulated data for all cases, which is consistent with the statistical results based on measured data in \cite{ZhangR17}. However, the variances for the EAoD and AAoA are quite different for different database.

\begin{figure}[tb]
	\centerline{\includegraphics[width=0.5\textwidth]{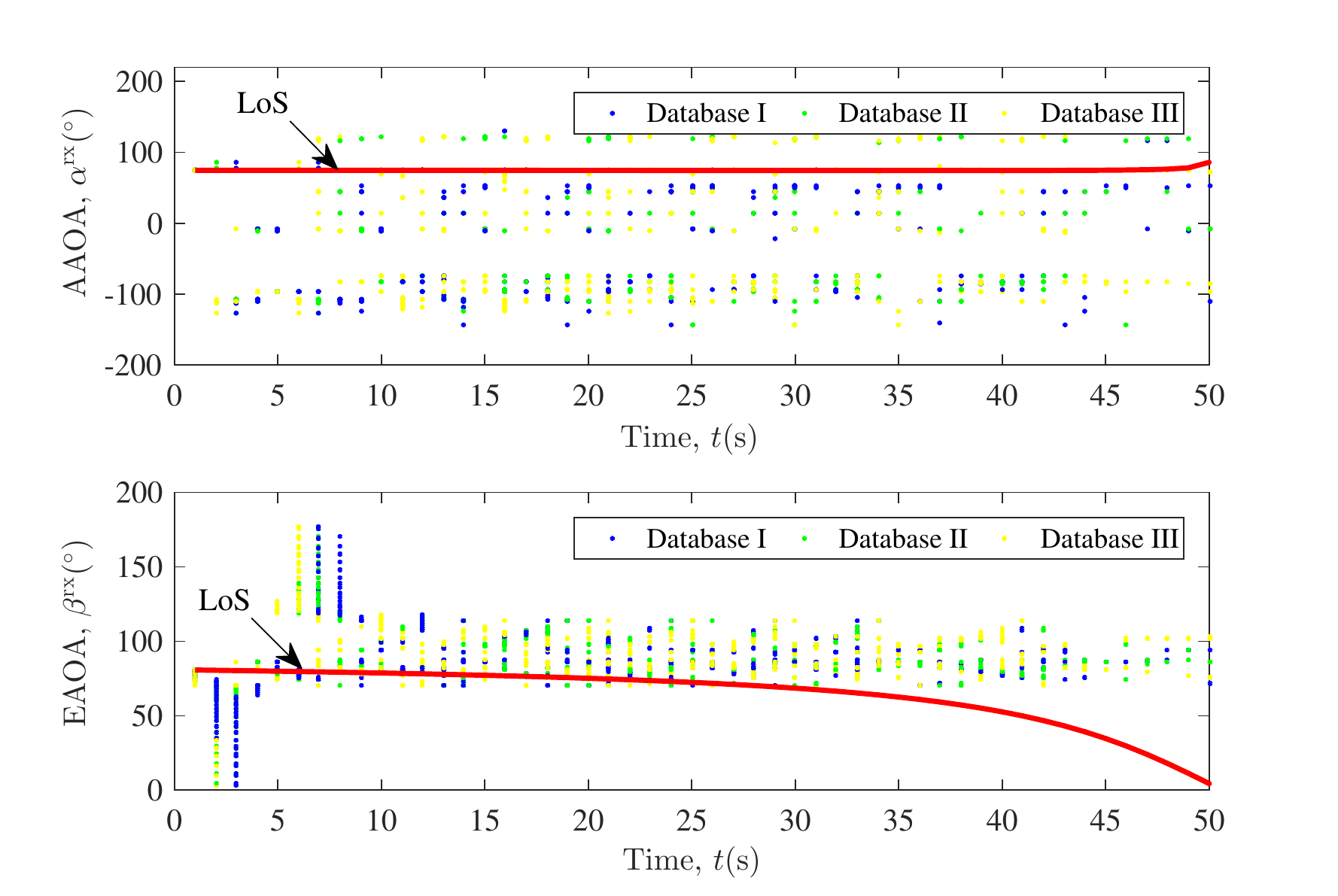}}
	\caption{Angle traces of AAoA and EAoA over time.}
	\label{fig:6}
\end{figure}




\section{Conclusions}
\par In this paper, we have developed a RT-based UAV-assisted mmWave channel model and studied on the effects of digital map accuracy on the channel parameters and properties. Since large city databases may take hours for the RT simulation, the reconstruction process for simplifying the original geometric database is demonstrated. Based on the reconstructed campus scene databases with different accuracies, the channel parameters of proposed model are obtained and analyzed. The simulation results have shown that three differently simplified databases have little effect on the channel. Therefore, we can select a proper reconstructed database according to the requirement to speed up the channel modeling process.

\begin{center}
ACKNOWLEDGMENTS
\end{center}
\par This work was supported by the National Key R\&D Program of China (No. 2018YF1801101), the National Key Scientific Instrument and Equipment Development Project (No. 61827801), the National Natural Science Foundation of China (No. 61960206006), the High Level Innovation and Entrepreneurial Research Team Program in Jiangsu, the High Level Innovation and Entrepreneurial Talent Introduction Program in Jiangsu, the Research Fund of National Mobile Communications Research Laboratory, Southeast University (No. 2020B01), the Fundamental Research Funds for the Central Universities (No. 2242019R30001), the EU H2020 RISE TESTBED2 project (No. 872172), the Huawei Cooperation Project, and the Open Foundation for Graduate Innovation of NUAA (No. KFJJ 20180408).
\begin{figure}[tb]
	\centerline{\includegraphics[width=0.5\textwidth]{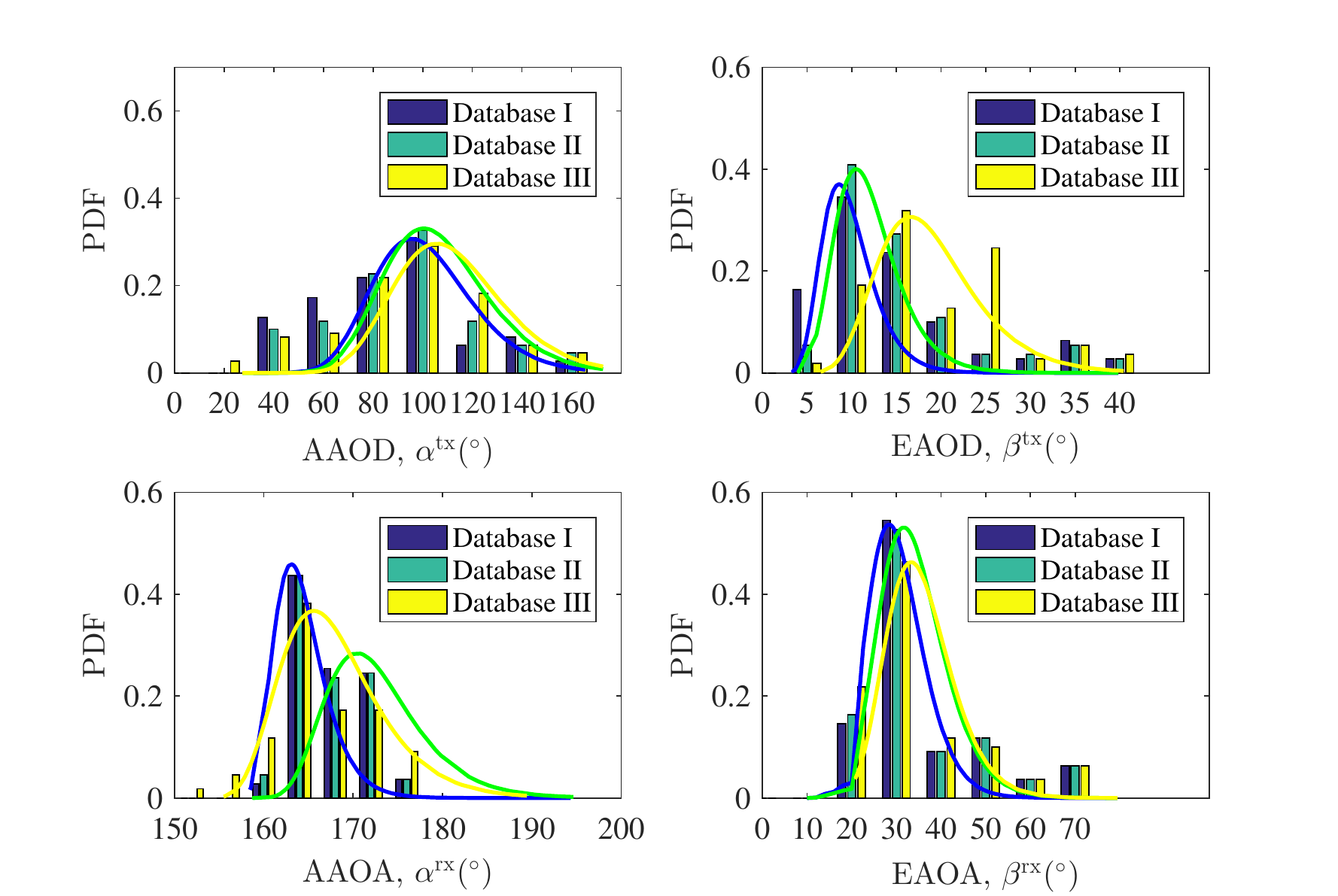}}
	\caption{Angle offset distributions of AAoD, EAoD, AAoA, and EAoA.}
	\label{fig:7}
\end{figure}

%








\end{document}